**Cockroaches adjust body and appendages to traverse cluttered large obstacles**

Yaqing Wang, Ratan Othayoth, *Chen Li

Department of Mechanical Engineering, Johns Hopkins University

*Corresponding author. Email: [redacted]

**Keywords:** terrestrial locomotion, locomotor transitions, potential energy barrier, terradynamics, kinematics

## Abstract

To traverse complex natural terrain, animals often transition between locomotor modes. It is well known that locomotor transitions can be induced by switching in neural control circuits or be driven by a need to minimize metabolic energetic cost. Recent work discovered that locomotor transitions in complex 3-D terrain cluttered with large obstacles can also emerge from physical interaction with the environment controlled by the nervous system. To traverse cluttered, stiff grass-like beams, the discoid cockroach often transitions from using a strenuous pitch mode to push across to using a less strenuous roll mode to maneuver through the gaps, during which a potential energy barrier must be overcome. Although previous robotic physical modeling demonstrated that kinetic energy fluctuation from body oscillation generated by leg propulsion can help overcome the barrier and facilitate this transition, the animal was observed to transition even when the barrier still exceeds kinetic energy fluctuation. Here, we further studied whether and how the cockroach makes active adjustments to facilitate this locomotor transition to traverse cluttered beams. We observed that the animal flexed its head and abdomen, reduced hind leg sprawl, and used both hind legs differentially during the pitch-to-roll transition, which were absent when running on a flat ground. Using a refined potential energy landscape with additional degrees of freedom modeling these adjustments, we found that head flexion did not substantially reduce the transition barrier, whereas the leg sprawl reduction did so dramatically. We discussed likely functions of the observed adjustments and suggested future directions.

## 1. Introduction

Animals locomotion emerges from direct physical interaction with the environment controlled by the nervous system, via both feedforward preflexes facilitated by morphology and feedback control modulated by sensing (Dickinson, 2000). To move across complex environments, animals often use and transition between multiple modes of locomotion (Alexander, 2006; Dickinson, 2000; Li et al., 2015; Lock et al., 2013; Low et al., 2015). Most terrestrial locomotion studies focused on how animals use neuromechanical control to generate or stabilize near-steady-state, single-mode locomotion (e.g., walking, running, (Blickhan and Full, 1993; Kuo, 2007)). Previous work explored how gait transitions result from changes in the rhythmic output of central pattern generators (Ijspeert, 2008), sensed information of the environment (Blaesing and Cruse, 2004; Ritzmann et al., 2012), or the need to minimize energetic cost over large spatiotemporal scales (Bramble and Lieberman, 2004; Shepard et al., 2013). Only very recently have we begun to understand that in complex 3-D terrain locomotor transitions can emerge from animals' direct physical interaction with the environment (Othayoth et al., 2020; Othayoth et al., 2021).

Our study was motivated by and builds upon this recent study, which concerns the discoid cockroach (*Blaberus discoidalis*) traversing a layer of cluttered grass-like beam obstacles (Othayoth et al., 2020). When encountering stiff beams, the animal often first pushes against the beams resulting in the body pitching body (the pitch mode), but then it often rolls its body into a gap between beams (the roll mode) to traverse. Potential energy landscape modeling revealed that the pitch and roll modes emerge as the system state is attracted to local minimum basins of an underlying potential energy landscape. Traversing stiff beams using the pitch mode is more strenuous than using the roll mode because it requires overcoming a higher potential energy barrier. Robotic physical modeling with feedforward control demonstrated that the system transitions from the pitch to the roll mode when the kinetic energy fluctuation from body oscillations (which result from cyclic leg propulsion) exceed a potential energy barrier between the pitch and roll basins. However, despite qualitatively similar overall findings, the animal's pitch-to-roll transition happens even when its body kinetic energy fluctuation is insufficient to overcome the barrier. This means that the animal must also be making active adjustments to facilitate the transition.

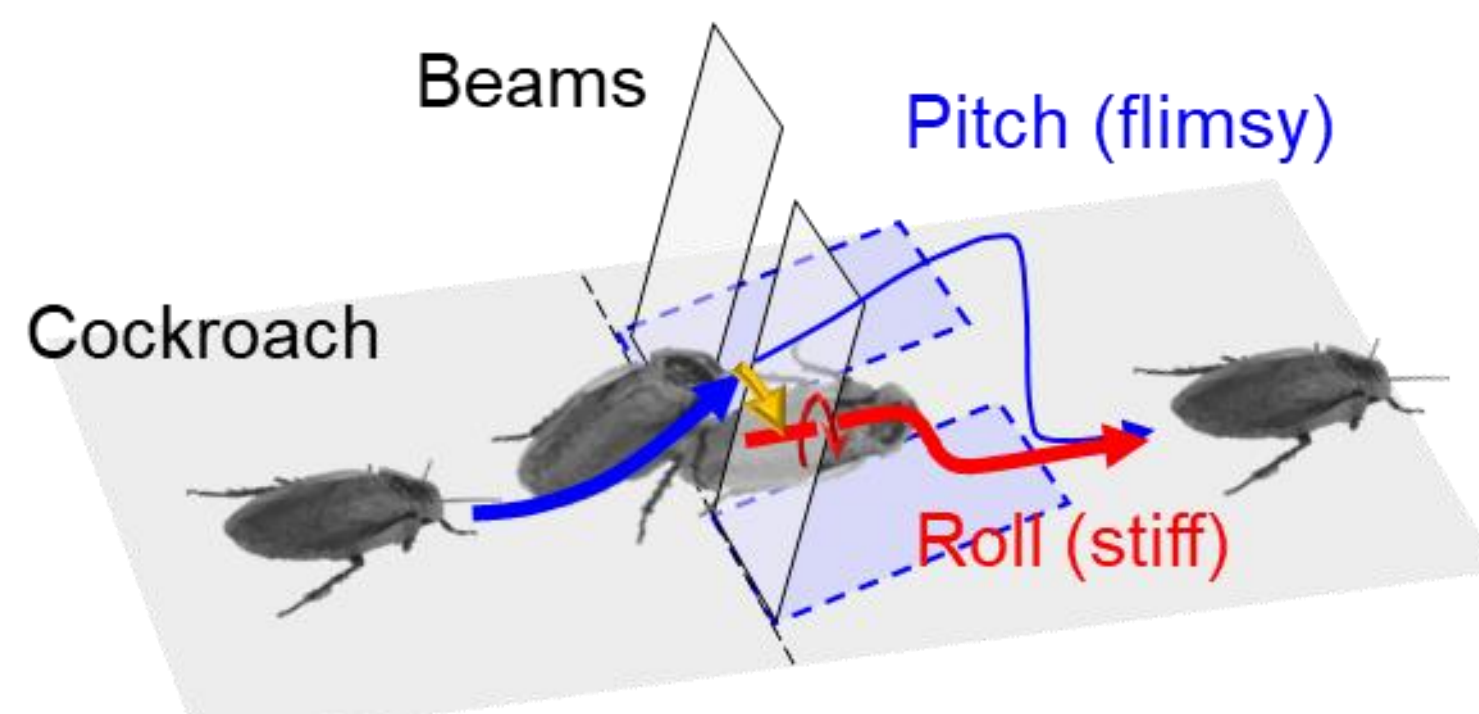

**Fig. 1. Cockroach transitioned from pitch to roll mode to traverse beams.** Adapted from (Othayoth et al., 2020).

Here we take the next step to further explore how the discoid cockroach make adjustments to facilitate pitch-to-roll transition. We challenged the animal to traverse a layer of stiff beams and used high speed imaging to measure detailed body and appendage kinematics. We discovered that the animal makes several adjustments: (1) Head flexion: the animal flexed its head up and down multiple times while interacting with the beams. (2) Abdomen flexion: the animal flexed its abdomen after the animal rolled into the beam gap. (3) Leg sprawl reduction: the animal spread its hind legs outward when pitching against the beams, but it tucked its hind legs inward when rolling into the beam gap. (4) Differential leg use: the animal extended one hind leg while retracting the other when rolling into the beam gap.

We hypothesized that the head flexion facilitates pitch-to-roll transition. Specifically: (1) When the animal transitioned from the pitch to the roll mode, active head flexion reduces the transition barrier and facilitates rolling into the gap. (2) When the animal body long has rolled into the gap, active head flexion facilitates it staying within the gap. We also hypothesized that the animal actively adjusts hind legs sprawl to facilitate pitch-to-roll transition. Specifically: (3) When pitching up against the beams, a large leg sprawl helps maintain the pitch-up state and reduce the tendency to roll. (4) When rolling into the gap, a small leg sprawl reduces the pitch-to-roll transition barrier.

To test hypotheses (1), (3), and (4) we used potential energy landscape modeling to analyze whether and how much the observed use of head flexion and leg sprawl reduction changed the potential energy barrier that must be overcome to transition from the pitch to the roll mode (which measures the difficulty of the transition). To test hypotheses (2), we analyzed whether and how much the observed use of head flexion changed the potential energy barrier that to transition from the rolled body being within the gap between the beams to being out of the gap and deflected sideways. We found that only leg sprawl reduction reduced the pitch-to-roll transition barrier. Finally, we discuss the likely functions of the observed body and appendage adjustments and suggest future directions.

## 2. Methods

### 2.1. Animals

We used three male *Blaberus discoidalis* cockroaches (Joe's BUGz LLC, Atlanta, GA, USA). Before the experiments, each animal was kept in a plastic container in a room with a controlled temperature of 22°C, moisture of 70%, and lighting on a 12h:12h light-dark cycle. Dry dog food (Purina Beneful, Largo, FL, USA) and water jelly made from water and polymer crystal (Tasty Worms Nutrition Inc., USA) were provided *ad libitum*. The animals weighed 2.7 ± 0.6 g (with marker items) and measured 5.3 ± 0.3 cm in length, 2.3 ± 0.1 cm in width, and 0.73 ± 0.08 cm in thickness. All data reported are means ± s.d. unless specified otherwise.

### 2.2. Obstacle track

For controlled, repeatable experiments, we constructed a testbed (Fig. 2A) similar to that in the previous study (Othayoth et al., 2020), with a layer of beam obstacles that consisted of seven beams. Each beam was 10 mm wide, 100 mm tall, and 0.8 mm thick. The lateral distance between two adjacent beams was 10 mm, and the lateral distance between the left/right-most beams and the walls was 5 mm. We used the same method to construct beam obstacles and characterize their stiffness as described in the previous study (Othayoth et al., 2020). The beams can only deflect about a hinge just above the ground. The beam torsional stiffness was $K = 2.5 \pm 0.4$ mN·m/rad (mean ± s.d. of 7 loading cycles), between the two most stiff beams in the previous study. We chose this high stiffness to induce a high pitch-to-roll transition probability (Othayoth et al., 2020) to increase experimental yield.

### 2.3. Imaging setup

Eight synchronized high-speed cameras (N5A-100, Adimec, Netherlands) recorded the experiment from different views: one from dorsal view, two from the side view, one from the oblique top view, and four from isometric views (Fig. 2A). All the cameras recorded at a frame rate of 100 Hz, a shutter time of 50 μs, and a resolution of 2592 × 2048 pixels. Even with eight cameras, we had to carefully tune camera positioning and orientation to achieve reliable tracking of the animal and beams (Sec. 2.5), because the animal had large 3-D body rotations (max absolute body yaw = 100°, max absolute pitch = 62°, max absolute roll = 98°, defined in Sec. 2.6) and markers were frequently occluded by the beams. Four work lamps (Coleman Cable, Waukegan, IL, USA) provided lighting from the top and side. During experiments, the ambient temperature around the arena was around 41°C.

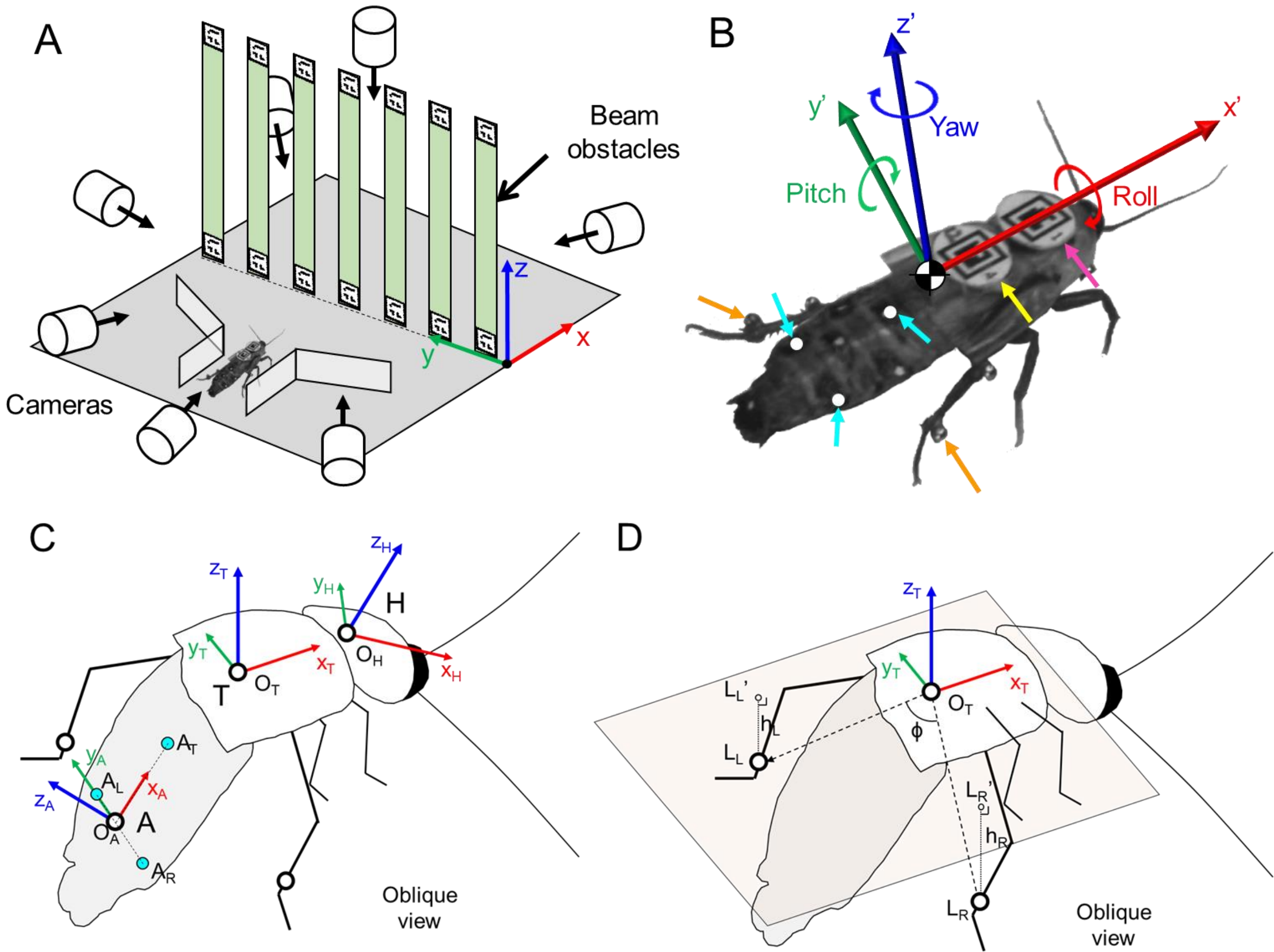

**Fig. 2. Experimental setup and cockroach schematic diagrams.** (A) Schematic of beam obstacle track and multi-camera imaging system. (B) Marker placement on animal and definition of thorax frame. Magenta: head marker. Yellow: thorax marker. Cyan: abdomen markers. Orange: leg markers. See body frame definition in Sec. 2.6. (C) Coordinate $X_TY_TZ_T$: thorax frame. Coordinate $X_HY_HZ_H$: head frame. Coordinate $X_AY_AZ_A$: abdomen frame. (D) Light-colored plane: thorax coronal plane. $L_L$': Left leg marker. $L_R$': Right leg marker. $\phi$: leg sprawl. $h_L$: left leg height. $h_R$: right leg height.

### 2.4. Experiment protocol

During each trial, we first placed the animal at the start of the track, covered it with a piece of cardboard, and let it settle down. We recorded the ambient temperature around the obstacle field and reset the beams to an upright pose. Then we started camera recording, lifted the cardboard to expose the animal to bright light, and prodded its abdomen with a tape-wrapped straw to induce running through the funnel towards the beams. After the animal traversed the beams, it entered a shelter of egg cartons (not shown) at the other end of the track. Then, camera recording was stopped, and videos were saved. The animal was allowed at least 3 minutes to rest after each trial. For each animal, we recorded 18-19 trials.

For each animal, we rejected the trials in which at least one of the following situations occurred: (1) The animal used at least one locomotor mode (Li et al., 2015) other than the pitch and roll modes (Othayoth et al., 2020) to traverse the beams. (2) The animal touched the arena wall in the roll phase (defined in Sec. 2.7). (3) At least one marker (BEEtags, white-outs, or beads) fell off. From the remaining trials, we selected the 12 trials with the least traversal time for each animal, with a total of 36 trials ($N = 3$, $n = 36$).

**2.5. Tracking and 3-D reconstruction**

To overcome the challenge in tracking from large 3-D body rotation and frequent occlusions, we used several types of markers (Fig. 2B). (1) We glued the animal's wings into a natural folded shape using hot glue and exposed the abdomen by trimming the posterior half of the wings. Then we used hot glue to attach a BEEtag (Crall et al., 2015) to the anterior half of the fixed wings covering the thorax as the thorax marker (Fig. 2B, yellow). (2) We used hot glue to attach a BEEtag onto the animals' pronotum as the head marker (Fig. 2B, magenta). (3) We used white-out to paint point markers on the dorsal surface of the abdomen as abdomen markers (Fig. 2B, cyan). (4) We used ultraviolet curing glue (Bondic, Aurora, Ontario, Canada) to attach two small, lightweight (12 mg each) aluminum beads (McMaster-Carr, Elmhurst, IL, USA) to each hind leg at two locations close to the femoral-tibial and tibia-tarsal joints as leg markers. (Fig. 2B, orange) (5) We attached BEEtags (Crall et al., 2015) to the top and the bottom of each beam's frontal side as beam markers (Fig. 2A).

Then, we tracked the markers on the animal and beams in each recorded video from all eight cameras. We tracked all the BEEtag markers automatically using a customed MATLAB code modified from the BEEtag code (Crall et al., 2015). To track the abdomen and leg markers efficiently, we used DeepLabCut (Mathis et al., 2018). For each camera view, we first manually digitized these markers in 10 trials, with 100 video frames from each camera view, and used these data as a training sample to train the neural network. After training, DeepLabCut tracked the markers in the videos. We then visually examined the sample tracking results, manually fixed obvious tracking errors, and re-trained the training sample. After several cycles of manual corrections and reinforcement learning, the DeepLabCut could automatically track markers with high accuracy. We visually checked the tracking result carefully and manually corrected the remaining tracking errors. Using this tracking method, we achieved a high maker tracking performance: the head, thorax, abdomen, and leg markers were all tracked in 100% of all the frames of all trials. We emphasize that, even with 8 cameras covering a large angular range and using DeepLabCut, tracking detailed kinematics in such a densely cluttered terrain is a laborious and time-consuming process, due to the large body rotation in 3-D and frequent occlusions of markers. In total, it took an experienced experimenter 20 hours of manual digitizing, 72 hours of automatic tracking, and 150 hours of manual correction to track 36 trials each averaging 280 frames with 8 camera views.

Finally, we reconstructed 3-D kinematics of all tracked markers using the Direct Linear Transformation method (DLTcal5) (Hedrick, 2008). To facilitate 3-D calibration, we built a calibration object with 60 BEEtag markers using Lego bricks (The Lego Group, Denmark).

### 2.6. Kinematics analysis

With the 3-D reconstruction of tracked markers, we quantified the motion of the animal's head, thorax, abdomen, and two hind legs. For simplicity, we laterally mirrored kinematic data of the trials in which the animal rolled to the left to become rolling to the right to simplify the analysis, considering lateral symmetry.

We first approximated the thorax frame (which is the body frame in (Othayoth et al., 2020)) and head frame using the BEEtags on them. To do this, we projected a model of the animal (Sec. 2.9) to a set of eight synchronized camera views, and adjusted its pose so that it visually matched the animal figure in the video. We checked this matching between the model and the animal figure in at least 5 other frames in the videos. Then we defined the thorax frame ($X_TY_TZ_T$) and head frame ($X_HY_HZ_H$) as the frame of the model, and used homogeneous transformation between the tags and the models to represent the spatial relationship between the tags and the animal thorax and head parts. For the abdomen frame ($X_AY_AZ_A$), we defined the origin ($O_A$) as the foot of the perpendicular from the top marker ($A_T$) to the segment of left ($A_L$) and right markers ($A_R$), defined the x-axis as the direction from the origin ($O_A$) pointing at the top marker ($A_T$), defined the y-axis as the direction from the origin ($O_A$) pointing at the left marker ($A_L$) (Fig. 2C). Thus, we obtained the head, thorax, and abdomen frames, each with 3-D position ($x$, $y$, $z$) and orientation (yaw $\alpha$, pitch $\beta$, roll $\gamma$, $Z$−$Y$'−$X$'' Tait-Bryan convention) (Fig. 2C).

We then calculated the following kinematic variables as a function of time in each trial. (1) Head flexion $\beta_h$, defined as the additive inverse of the pitch of the head frame in the thorax frame. (2) Abdomen flexion $\beta_a$: the pitch of the abdomen frame in the thorax frame. (3) Leg sprawl $\phi$: the angle between the two vectors from the thorax frame origin ($O_B$) to the leg markers ($L_L$, $L_R$) (Fig. 2D). (4) Leg height difference $\Delta h$: The leg height of the left hind leg ($h_L$) minus the leg height of the right hind leg ($h_R$) (after mirrored). Hind leg height was defined as the distance of the leg marker from the thorax coronal plane (Fig. 2D). All the Equations are summarized below.

The rotation matrix of the thorax ($R_B$), head ($R_H$), and abdomen ($R_A$) frames in the world frame:

$$R = \begin{bmatrix} c_\alpha c_\beta & c_\alpha s_\beta s_\gamma - s_\alpha c_\gamma & c_\alpha s_\beta c_\gamma + s_\alpha s_\gamma \\ c_\alpha s_\beta & c_\alpha s_\beta s_\gamma + s_\alpha c_\gamma & c_\alpha s_\beta c_\gamma - s_\alpha s_\gamma \\ -s_\beta & c_\beta s_\gamma & c_\beta c_\gamma \end{bmatrix},$$

Which $s_\times$ and $c_\times$ are abbreviations for sine and cosine terms, respectively. $\alpha$, $\beta$, and $\gamma$ are the Euler angles.

The rotation matrix of head and abdomen frames in the thorax frame:

$$R_{TH} = R_T^{T} R_H,$$

$$R_{TA} = R_T^{T} R_A,$$

The head flexion and abdomen flexion:

$$\beta_h = \beta(R_{TH}),$$

$$\beta_a = -\beta(R_{TA}),$$

Where $\beta(\cdot)$ means to obtain the pitch of a rotational matrix:

$$\beta(R) = \text{atan2}(\sqrt{r_{31}^2 + r_{32}^2}, r_{33}),$$

Where $r_{ij}$ is the i[th] row and j[th] element in the matrix $R$.

Leg sprawl:

$$\phi = \cos^{-1} [(\overrightarrow{O_T L_L} \cdot \overrightarrow{O_T L_R}) / (\| \overrightarrow{O_T L_L} \| \cdot \| \overrightarrow{O_T L_R} \|)]$$

Leg height difference:

$$\Delta h = Z(L_L) - Z(L_R),$$

where $Z(\cdot)$ means to obtain the $z$ coordinate of a point.

To obtain average kinetic energy fluctuation, we first calculated the time average of the total kinetic energy due to translational and rotational velocity components other than in the forward motion (Othayoth et al., 2020) in the explore + pitch and roll phases for each trial, then averaged the means of all trials.

All kinematic analyses were performed using MATLAB (MathWorks, Inc., MA, USA).

### 2.7. Definition of traversal phases

To compare the animal's motion in different stages of the traversal, we divided each trial into five distinct phases:

(1) Approach (gray): From when the animal ran into the camera view to when it collided with the beams.

(2) Explore + Pitch (blue): From when the animal collided with the beams to when it started the final, successful roll attempt. Because the animal sometimes attempted to roll its body more than once, here we separated this phase and the next phase with the start of the last, successful attempt. The start of the attempt was defined as the instance when the animal's body roll changed sign from negative to positive.

(3) Roll (red): From when the animal started the final, successful body roll attempt to when body roll was maximal.

(4) Land (orange): From when body roll was maximal to when the animal landed with all its six legs had touched the ground again. In all the following analyses in this article, this phase was ignored.

(5) Depart (green): From when the animal landed with all its six legs touching the ground to when it exited the camera view.

### 2.8. Statistics

We aligned the time axis of each trial by setting the beginning of the roll phase as $t = 0$. For each time instance, we calculated the mean and standard deviation of each variable across all the trials. To compare across traversal phases, we first averaged all kinematic variables (i.e., body roll $\gamma$, body pitch $\beta$, head flexion $\beta_h$, abdomen flexion $\beta_a$, leg sprawl $\phi$, and leg height difference $\Delta h$) over time in each phase, except for the head flexion $\beta_h$. Because the standard deviation of the head flexion $\beta_h$ can better reflect the animal's repeated head flexion than the average does, we calculated the standard deviation of the head flexion $\beta_h$ over time in each phase. All average data are reported as mean ± s.d. These analyses were performed using MATLAB (MathWorks, Inc., MA, USA).

Using these data, for each pair of phases, we performed a mixed-design ANOVA, with the phase as a fixed factor, and the individual as a random factor to account for individual variability. These statistical tests were performed using JMP 16 (SAS Institute Inc., NC, USA).

### 2.9. Potential energy landscape model definition

In the previous study, to generate the potential energy landscape, the animal body was modeled as a single rigid body (Othayoth et al., 2020). Here, to further study how the active adjustments facilitated the beam traversal, we refined the animal body model to consist of a head, a thorax, and an abdomen (Fig. 3A). Antennas, front and middle legs, and other body parts were neglected. The hind legs were neglected when studying the usefulness of head flexion (Sec. 2.12), considering that they are not in contact with beams during head-beam interaction. The hind legs were added when studying the use of leg adjustments (Sec. 2.13). The thorax was modeled as a half ellipsoid (Fig. 3A, orange, length: 19.9 mm, width: 27.6 mm, thickness: 6.4 mm). The head was modeled as a massless ellipsoid-like rigid body (Fig. 3A, red, length: 9.1 mm, width:14.4 mm, thickness: 6.0 mm). The abdomen was modeled as half of an ellipsoid (Fig. 3A, yellow, length: 30.5 mm, width: 20.6 mm, thickness: 6.5 mm). All the dimensions above were the averages of the measured animal dimensions. When added, the hind legs were modeled as rigid rods with one side fixed to the thorax center (Fig. S2B), because we only tracked the tibia-tarsal joint of the legs. The length of each leg was 27 mm (average of maximal leg length is 27 ± 2 mm in the explore + pitch and roll phases over all trials). For simplicity, we assumed that the head and abdomen could each only flex about a lateral axis fixed to the thorax (i.e., only the pitch degree of freedom was allowed). We set the body center of mass at the middle of the rotation axes between the thorax and the abdomen, which is a reasonable approximation (Kram et al., 1997). We intentionally designed overlapping between the thorax and the head or abdomen (Fig. 3B) to reduce unrealistic concavity between these segments.

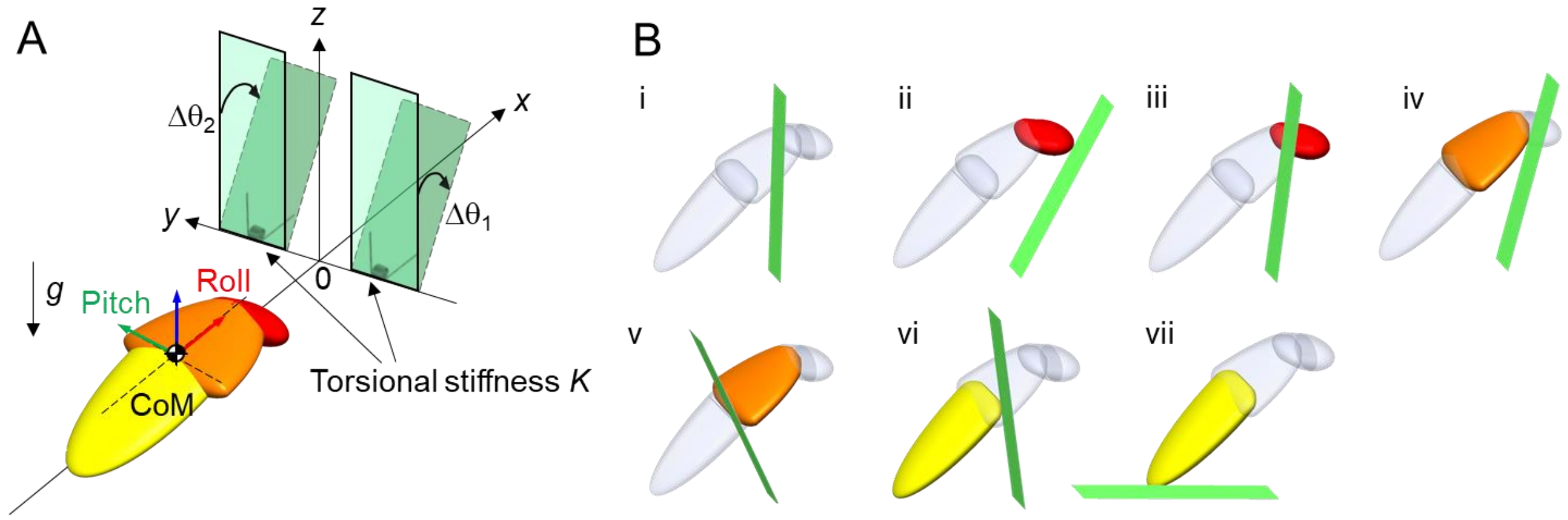

**Fig. 3. Modeling for potential energy landscape approach.** (**A**) Model of animal and beams and definition of variables. (**B**) Schematic examples to choose between multiple possibilities of beam deflection.

The beams were modeled as rigid rectangular plates attached to the ground with Hookean torsional joints at the bottom, and their orientations without animal interaction were set to vertical (Fig. 3A, green solid). In the previous study, the beams were only allowed to deflect forward, and the largest possible deflection angle were always selected (Othayoth et al., 2020). This resulted in overestimated beam deflection. In particular, when the cockroach had already traversed the beams using the roll mode, the estimated beam deflections were still calculated as if the beams blocked in front of the cockroach when the animal used a pitch mode. Here, to refine the model, we allowed the beams to deflect either forward or backward during the interaction, and we determined each beam's deflection ($\Delta\theta_1, \Delta\theta_2$) as the angle with minimal absolute value at which the beam did not overlap with any part of the animal. This revised protocol ensured that, when the cockroach is sufficiently far away from the beam, either not having entered the beam area or having already traversed, beam deflection is zero; when the animal is interacting but has not traversed the beams, the beams deflect forward; when the animal has traversed the beams, the beams deflect backward. Note that this revised protocol does not affect the transition barrier analysis, because the pitch-to-roll transition happened when the animal body was only beginning to enter the gap(average body $x$ = −13.6 ± 4.4 mm when pitch-to-roll transition happened over all trials), in that case, both protocol gives the same result.

Below we give an example of how to determine beam deflection (Fig. 3B). We first identified seven possible deflection angles, i.e., 0° deflection (Fig. 3B, i), deflections where the beam is tangential to the head in the front or back (Fig. 3B, ii, iii), to the thorax in the front or back (Fig. 3B, iv, v), or to the abdomen in the front or back (Fig. 3B, vi, vii). Sometimes, there are no deflections where the beam is tangential to any body part; in that case, the two possible deflection angles where the beam is tangential to this part are set to be 0°. Then, we rejected the beam deflections where the beam overlaps with any body part (Fig. 3B, i, iii, iv, v, vi), and finally selected the deflection angle with minimum absolute value (Fig. 3B, ii is selected, Fig. 3B, vii is rejected). Beam deflection calculated from this method better matched experimental measurements than in the previous study (Othayoth et al., 2020), reducing the error from 15° ± 32° to −1° ± 13° ($P < 0.001$, repeated-measures ANOVA).

The potential energy of the system is the sum of animal and beam gravitational potential energy and beam elastic energy:

$$E_{potential} = m_{cockroach} g z + \frac{1}{2} m_{beam} g L (\cos \Delta\theta_1 + \cos \Delta\theta_2) + \frac{1}{2} K (\Delta\theta_1^2 + \Delta\theta_2^2), \quad [1]$$

where $m_{cockroach}$ is the animal mass, $g$ is gravitational acceleration, $z$ is the body center of mass height from the ground, $m_{beam}$ is the beam mass, $L$ is the beam length, $K$ is the beam torsional stiffness, and $\Delta\theta_1$ and $\Delta\theta_2$ are beam deflection from vertical. Given the constraints above, it was fully determined by the animal's position, orientation, and head and abdomen flexion and does not depend on the trajectory (i.e., determined by configuration with no history dependence). This is crucial for the application of potential energy landscape approach, because it simplifies the problem to be within a finite number of dimensions and further makes variation of variables practical.

**2.10. Potential energy landscape model generation**

We generated the potential energy landscape in a similar way as in the previous study (Othayoth et al., 2020). The model system has eight degrees of freedom, including the animal position (forward $x$, lateral $y$, vertical $z$) and orientation (yaw $\alpha$, pitch $\beta$, roll $\gamma$) of thorax and head ($\beta_h$) and abdomen ($\beta_a$) flexion. So, the potential energy of the system should be a function of these eight independent variables:

$$E_{potential} = E_{potential}\ (x, y, z, \alpha, \beta, \gamma, \beta_h, \beta_a) \quad [2]$$

We calculated the potential energy landscape over the 8-D space by varying these eight variables, their ranges, and increments are listed in Tab. 1. We did not vary the abdomen flexion $\beta_a$ for two reasons. First, the head and leg adjustments likely facilitated transition to the roll mode, whereas the abdomen which interacted with the beams after the body had already rolled into the gap and likely contributed less to this transition. Second, it is computationally costly to add one more dimension to our potential energy landscape calculations. The first 7 dimensions that we varied systematically already took 3 weeks of computation on a 32-core 2.93 GHz workstation. It would take ~20 times more (~a year) if we varied abdomen flexion like head flexion. To simplify landscape analysis, we focused on two cross-sections of the full 8-dimension landscape over a few dimensions of interest by collapsing less relevant dimensions. We first collapsed the landscape along the $z$ dimension. For each combination of the other 7 variables, potential energy is a function of $z$. We varied $z$ from $z_{min}$ (when the body touched the ground) to $z_{min}$ + 15 mm and chose the $z$ value for which potential energy is minimal, assuming that the unstable (due to self-propulsion) system was attracted to the minimum. This method was different from that used in the previous study (Othayoth et al., 2020), where the animal's lowest point was constrained to always touch the ground (i.e., ground contract constraint). The body $z$ obtained from this refined method better matched observations: with the ground contact constraint, to reach the average measured $z$ when the animal interacted with the beams ($x = -9$ to

−3 mm, Fig. S2B) would require the animal to pitch up by an average of 20°, much greater than the observation (< 10°, Fig. S2D) which was only possible without the ground contact constraint.

Next, we collapsed the landscape along other less relevant dimensions by analyzing landscape cross-sections that follow the average animal trajectory. To extract an average animal trajectory as a function of forward position $x$, we first discretized $x$ within [−26, 33] mm into 296 bins each spanning 0.2 mm. Between each two adjacent time steps in each trial, we checked whether the animal passed any of these $x$ bins. For each bin where this occurred, we determined the values of kinematic variables other than $x$ using linear interpolation over $x$ and recorded them under this bin. Finally, we averaged these recorded variables for each $x$ bin, the evolution of which over $x$ gave the average animal trajectory (Fig. S2). For the yaw cross-section analysis below, we always kept body $y$, body pitch $\beta$, and roll $\gamma$ to follow the average trajectory. For the pitch-roll cross-section analysis, we always kept body $y$ and yaw $\alpha$ to follow the average trajectory. For both analyses, we constrained the abdomen pitch $\beta_a$ fixed at 7° (temporal average of abdomen flexion was $\beta_a = 7° \pm 4°$ in the approach phase over all trials). Head flexion $\beta_h$ was a variable in Sec. 2.12 and a was set to follow the average trajectory in Sec. 2.13. See Tab. 1 for a summary of the ranges and increment of parameter variation and dimension collapsing protocol.

To study how head and leg adjustments affected pitch-to-roll transition (Sec. 2.12, 2.13), we extracted a body yaw cross-section of the landscape, where potential energy is a function of body yaw $\alpha$. To study if the head flexion helped the animal stay within the gap after the animal rolled (Sec. 2.12), we extracted a pitch-roll cross-section of the landscape, where potential energy is a function of body pitch $\beta$ and roll $\gamma$.

**Table 1. Ranges and increment of landscape variation and dimension collapse protocol.** $z_{min}$ is the $z$ where the lowest point of the body touches the ground.

| Variable | Unit | Min | Max | Increment | Dimension collapsing protocol | |
|---|---|---|---|---|---|---|
| | | | | | Pitch-roll cross-section | Yaw cross-section |
| Forward position $x$ | mm | −26 | 33 | 0.2 | NC | NC |
| Lateral position $y$ | mm | −3 | 3 | 1 | FAT | FAT |

| Vertical position $z$ | mm | $z_{min}$ | $z_{min}$ + 15 | 1 | MPE | MPE |
|---|---|---|---|---|---|---|
| Yaw $\alpha$ | ° | −90 | 90 | 5 | FAT | NC |
| Pitch $\beta$ | ° | −90 | 90 | 2 | NC | FAT |
| Roll $\gamma$ | ° | −180 | 180 | 2 | NC | FAT |
| Head flexion $\beta_h$ | ° | −25 | 65 | 5 | NC or FAT | NC |
| Abdomen flexion $\beta_a$ | ° | 7 | 7 | - | - | - |

NC: not collapsed. FAT: follow average trajectory. MPE: minimize potential energy.

### 2.11. Quantifying difficulty of transition using potential energy barrier

First say that you are analyzing the transition barrier from the pitch to the roll mode for hypotheses (1) and (2), and the other barrier

To quantify the difficulty to transition between two modes (i.e., from the pitch to the roll mode, from the roll to the deflect mode), we generated the relevant cross-section of the landscape, identified the basins corresponding to these modes, and calculated the transition barrier.

First, we looped through all points on the landscape to identify the local minimum basins corresponding to the locomotor modes. For the pitch-roll cross-section, we located the pitch minimum (Fig. 4B, blue points on the cross-sections) with a finite body pitch and zero body roll (Fig. 4A, i) and the roll minimum (Fig. 4B, red points on the cross-sections), with a body pitch near 0° and a body roll around 90° (Fig. 4A, ii). When no roll minimum existed (e.g., when the animal was far from the beams, Fig. 4B, i), we defined (pitch, roll) = (0°, 90°) as the roll minimum. For the yaw cross-section, we located the between-gap minimum (Fig. 4C, blue points on the cross-sections) with a body yaw around 0° between the gap (Fig. 4A, ii) and the deflect minima (Fig. 4B, red points on the cross-sections) with the body deflected towards the left or right with a body yaw around ±90° (Fig. 4A iii, iii'). Note that the roll basin on the pitch-roll cross-section and between-gap basin on the yaw cross-section correspond to the same body configuration (Fig. 4A, ii).

To transition from one mode to another, the animal had to overcome a potential energy barrier (i.e., transition barrier) on the landscape cross-section. A higher transition barrier means that it is more difficult to transition. We can measure whether and how the transition barrier changed with an observed adjustment (i.e., head flexion, leg sprawl reduction) to evaluate whether it facilitated or hindered a transition. We used breadth-first search (BFS) to calculate transition barriers of the transitions from one basin to another. Breadth-first search is a computational algorithm for traversing graph data structure (Cormen et al., 2009). We fed the gridded potential energy landscape cross-section as the graph data, with the starting minimum as the start point, the destination minimum as the goal point, and the highest potential energy on the traversing route as the cost function. The algorithm gave a route from the starting minimum to the destination minimum that crossed the lowest energy barrier (Fig. 4B, C, green curve). We defined the point with the highest energy on this route as the saddle point (Fig. 4B, C, orange point), which was usually a local maximum on a landscape or a true saddle point on a landscape. We defined the transition barrier as the potential energy increasing from the starting minimum to the saddle.

Intuitively, the breadth-first search approach resembled to extrude water at the start point and track the expansion of the water-covering area. The potential energy landscape resembled an uneven surface, and each basin on the surface corresponded with a locomotor mode, such as pitch and roll modes (Othayoth et al., 2020). Enlarging the cost resembled extruding water in the starting basin and increasing the water level. As the water level increased, there was a moment that the water level was high enough to overwhelm a gap between two basins, and the water flowed via this gap into the destination basin. The current water level resembled the potential energy barrier, and the water flow resembled the rest searching sequence in the breadth-first search approach.

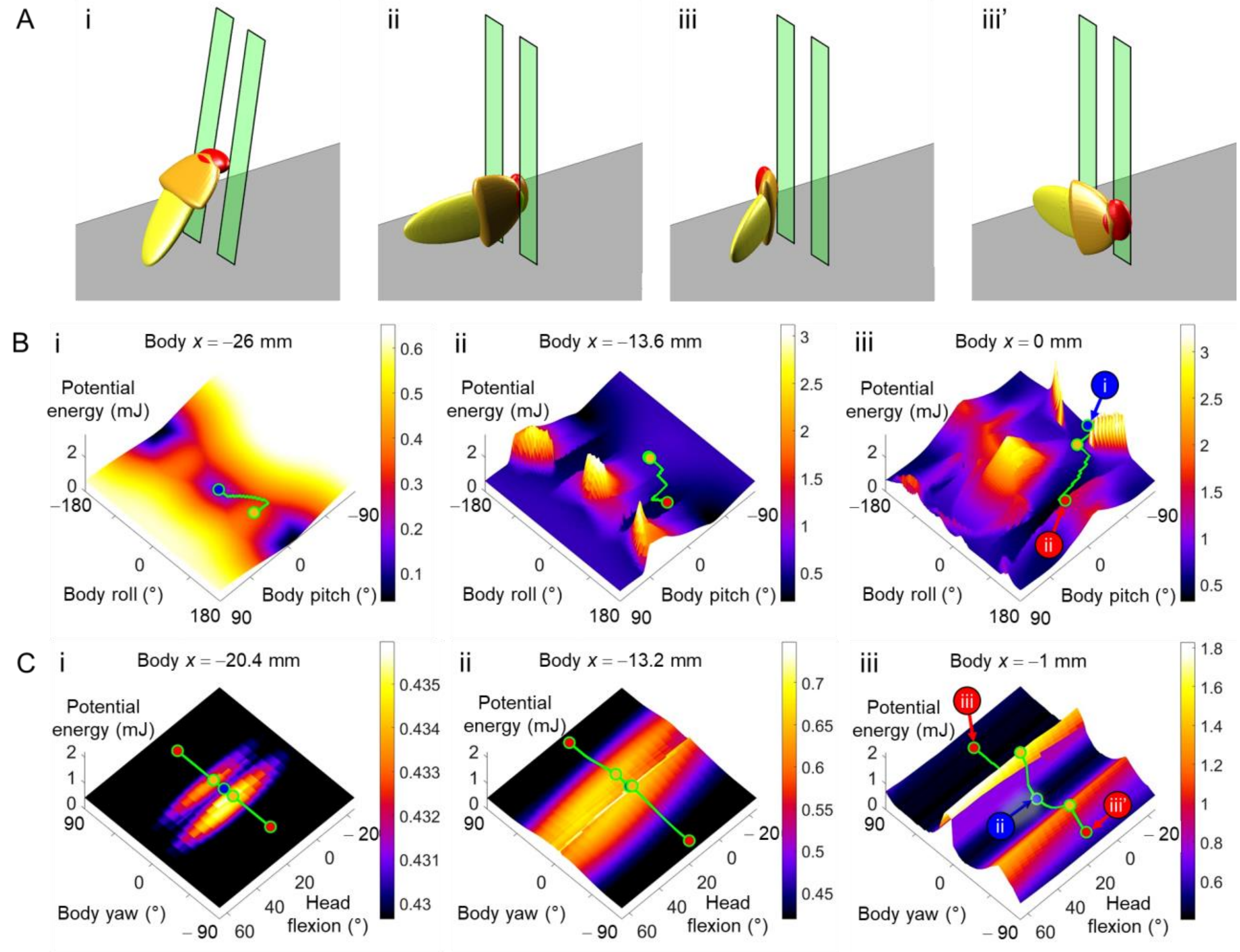

**Fig. 4. Evolution of potential energy landscape.** (**A**) Demonstration of locomotor modes. (i) Pitch mode. (ii) Roll mode, also between-gap mode. (iii) Deflect mode, to the left. (iii’) Deflect mode, to the right. (**B**) Pitch-roll cross-section at different *x* for analyzing pitch-to-roll transition barrier. Blue points are pitch local minima. Red points are roll local minima. Orange points are saddle points. Green curves: imaginary routes (see Sec. 2.10). The example shown is at head flexion $\beta_h = 15°$ and without hind legs. (i) $x = -26$ mm. Roll minimum and saddle point overlap. (ii) $x = -13.6$ mm. Pitch minimum and saddle point overlap. (iii) $x = 0$ mm. (**C**) Yaw cross-sections stacked along head flexion dimension for different *x* for analyzing between-gap-to-deflect transition barrier. Blue points are between-gap local minima, the one with large body yaw is deflecting to the left. Red points are deflect local minima. Orange points are saddle points. Green curves:

imaginary routes. Only routes at head flexion $\beta_h$ = 15° are marked. (i) $x$ = −20.4 mm. (ii) $x$ = −13.2 mm. (iii) $x$ = −1 mm.

### 2.12. Test usefulness of head flexion

To test hypothesis (1), we analyzed whether head flexion in the range observed reduced the pitch-to-roll transition barrier compared to if the animal simply held its head in the typical orientation (average head flexion during running on flat ground). We varied head flexions within [–25°, 65°] (covering the observed head flexion range of [–24°, 64°] over all the trials) with an increment of 5° (Fig. S2A), calculated the pitch-to-roll transition barrier as a function of $x$, and compared the transition barrier at each head flexion with that at head flexion $\beta_h$ = 15°, which represents the case without active head flexion (temporal average of head flexion was $\beta_h$ = 15° ± 4° in the approach phase over all trials).

To test hypothesis (2), we analyzed whether the head flexion in the range observed increased the between-gap-to-deflect transition barrier compared to without active head flexion. We varied the head flexion like above, calculated the between-gap-to-deflect transition barrier as a function of $x$, and compared the transition barrier at each head flexion with that at head flexion $\beta_h$ = 15°.

### 2.13. Test usefulness of leg sprawl reduction

To test hypotheses (3) and (4), we analyzed whether the leg sprawl changes affected the between-gap-to-deflect transition barrier. We added two hind legs (length = 27 mm) into the model symmetrically to the left and right sides of the body, with a leg height of −5 mm (temporal average of leg height was −5 ± 3 mm in the explore + pitch phase (Sec. 3.1) over all trials), and varied leg sprawl angles within [0°, 159°] with an increment around 10° (Fig. S2B). We compared the transition barrier at each leg sprawl angle with that at leg sprawl $\phi$ = 143° (average maximal leg sprawl was $\phi$ = 140° ± 18° in the explore + pitch phase over all trials) and $\phi$ = 39° (average minimal leg sprawl was $\phi$ = 40° ± 18° in the roll phase over all trials).

## 3. Results

### 3.1. Animal uses complex motion to transition from pitch to roll mode

After running with an alternating tripod gait and collided with the beams (approach phase, Fig. 5, gray), the animal traversed the beam obstacles with complex body, head, and leg motions. In the explore + pitch phase (Fig. 5, blue), the animal often moved along the beam layer (*y*-direction) and turned left or right to search around the beams, pitched up its body against the beams, flexed its head up and down, and rubbed its pronotum against the beams, sometimes pushed the beam using its fore or middle legs, and swept its antennas in the gaps. In the roll phase (Fig. 5, red), the animal rolled its body into the gap, flexed its head repeatedly, sometimes flexed and twisted its abdomen, and struggling its legs to try to push the back surface of the beam. In the land phase (Fig. 5, orange), the animal passed through the beams in the roll mode, and resumed an upright body orientation. In the depart phase (Fig. 5, green), the animal ran away in an alternating-tripod gait. These observations were consistent with those in the previous study (Othayoth et al., 2020). The average animal trajectory as a function of forwarding position *x* is shown in (Fig. S2).

Similar to the previous study, body oscillation was observed (Othayoth et al., 2020). The average kinetic energy fluctuation was 0.01 ± 0.01 mJ. This is smaller than that in the previous study (0.02 ± 0.01 mJ). We speculate that this is caused by the additional modifications on the animal (wing trimming, adding one more tag on the pronotum and beads on hind legs) that slightly resisted the animal motion.

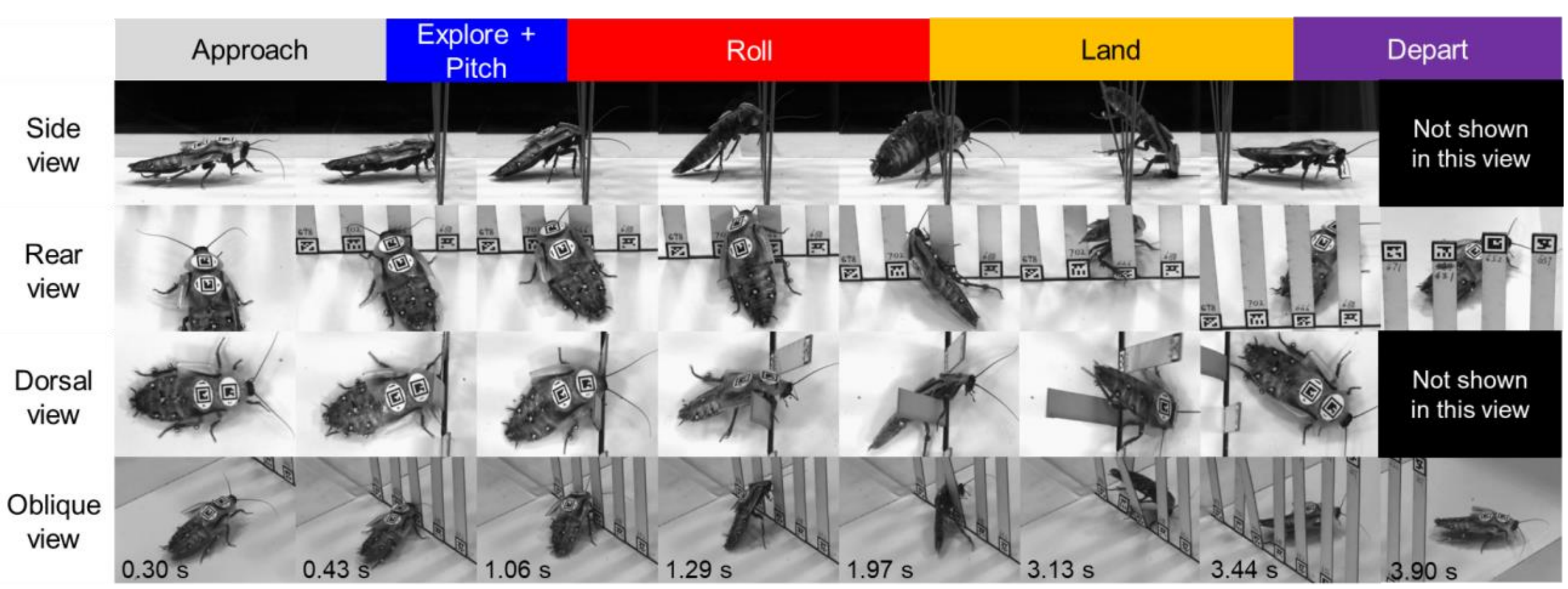

**Fig. 5. Five phases of traversal.**

### 3.2. Body rotations

The animal's body pitched up in the explore + pitch phase, and rolled while pitched down in the roll phase. In the roll phase, the animal rolled its body significantly (maximum roll $\gamma = 81° \pm 10°$) (Fig. 6A). Before the animal rolled, its body pitch increased (maximum pitch $\beta = 36° \pm 9°$). Then, as the animal rolled, body pitch decreased (Fig. 6C). The average body roll in the roll phase was higher than in the other three phases ($P < 0.0001$, repeated-measures ANOVA, Fig. 6B). Average body pitch in the explore + pitch phase was higher than in the approach and depart phases ($P < 0.0001$, repeated-measures ANOVA, Fig. 6D) and was slightly higher in the roll phase than in the approach and depart phases ($P < 0.05$, repeated-measures ANOVA, Fig. 6D).

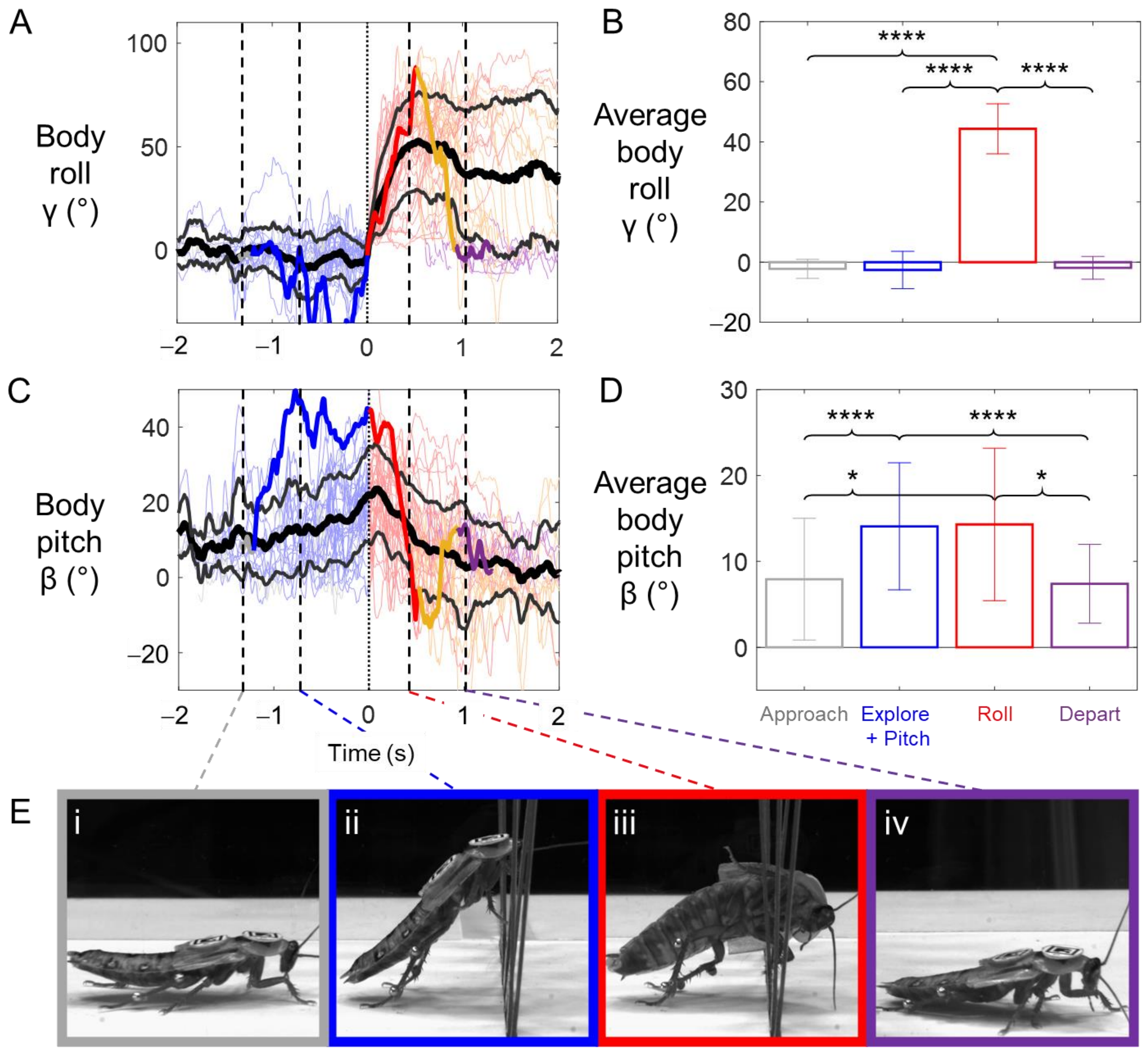

**Fig. 6. Body rotations.** (**A**) Body roll as a function of time. (**B**) Average body roll in different phases. (**C**) Body pitch as a function of time. (**D**) Average body pitch in different phases. Colors are for five phases defined in Fig. 5. In (**A**, **C**), thin colored curves are individual trials; thick and thin black curves are mean ± s.d. across all trials, and thick colored curve is an example individual trial. In (B, D), bars and error bars are means ± s.d. of the temporal averages of all trials in (A, C) for each phase. (**E**) Representative snapshots for each phase.

### 3.3. Head flexion

During the explore + pitch and roll phase, the animal sometimes flexed its head up and down dynamically and sometimes flexed its head down and held it statically. (Fig. 7A, Movie 1). The standard deviation of the head flexion was higher in the explore + pitch and roll phases than in the approach and depart phases ($P < 0.0001$, repeated-measures ANOVA, Fig. 7B) and was a slightly higher in the depart phase than in the approach phase ($P < 0.01$, repeated-measures ANOVA) (Fig. 7B).

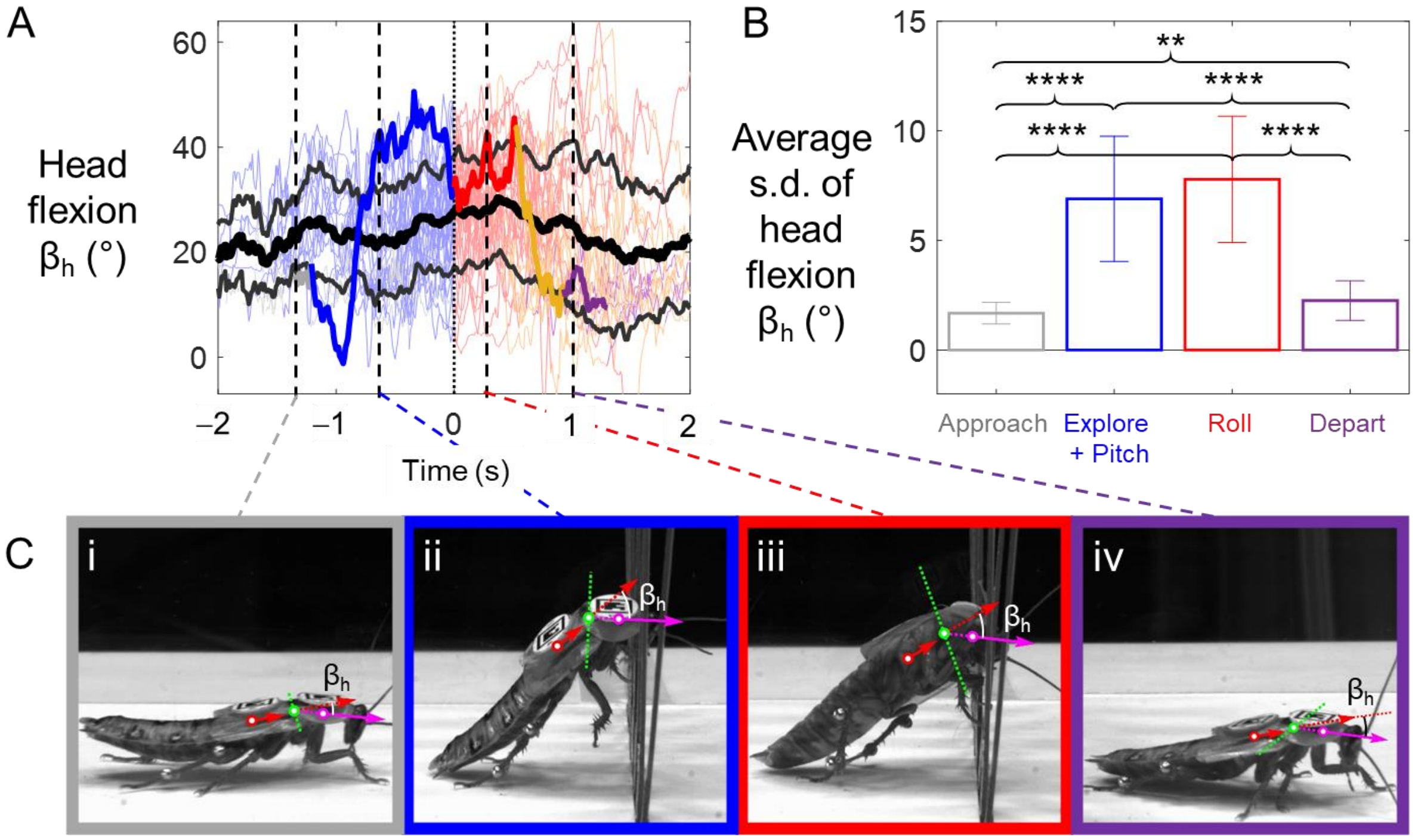

**Fig. 7. Head flexion.** (**A**) Head flexion as a function of time. Thin colored curves are individual trials; thick and thin black curves are mean ± s.d. across all trials. (**B**) Average standard deviation of head flexion in different phases. Bars and error bars are means ± s.d. of the temporal standard deviations of data in (A) for each phase. (**C**) Representative snapshots for each phase.

### 3.4. Abdomen flexion

The animal flexed its abdomen at a larger amplitude in the roll phase than in the other phases (Movie 2). In the approach and explore + pitch phase, the animal's abdomen only flexed down slightly (average abdomen flexion $\beta_a = 7° \pm 4°$). As the body began to roll, abdomen flexion increased (maximum abdomen flexion $\beta_a = 37° \pm 12°$) (Fig. 8A). Average abdomen flexion was higher in the roll phase than in the three other phases ($P < 0.0001$, repeated-measures ANOVA, Fig. 8B) and was slightly higher in the explore + pitch phase than in the approach phase ($P < 0.05$, repeated-measures ANOVA, Fig. 8B).

**Fig. 8. Abdomen flexion statistics.** (**A**) Abdomen flexion as a function of time. Thin colored curves are individual trials; thick and thin black curves are mean ± s.d. across all trials. (**B**) Average abdomen flexion angle in different phases. Bars and error bars are means ± s.d. from temporal averaging of data in (A) for each phase. (**C**) Representative snapshots for each phase.

### 3.5. Leg adjustments

The animal adjusted its hind legs in the roll phase in two ways (Movie 3). First, it tucked its hind legs inward in the roll phase. During the explore + pitch phase, the animal sprawled its hind legs outward (average leg sprawl = 115° ± 16°). As rolling began, the animal tucked its hind legs inward (average leg sprawl = 84° ± 15°) (Fig. 9A). Average leg sprawl was larger in the explore + pitch phase than in the other three phases ($P < 0.0001$, repeated-measures ANOVA, Fig. 9B) and smaller in the roll phase than in the other three phases ($P < 0.0001$, repeated-measures ANOVA, Fig. 9B). Average leg sprawl was slightly different in the depart phase from the approach or the explore + pitch phase ($P < 0.05$, repeated-measures ANOVA, Fig. 9B).

In addition, the animal extended one hind leg while retracting the other in the roll phase. In the explore + pitch phase, both hind legs had similar heights (average leg height difference = 0 ± 2 mm. In the roll phase, one hind leg extended (maximum leg heights = 19 ± 3 mm), appearing to push against the ground, while the other retracted (minimum leg heights = − 4 ± 2 mm), which increased the leg height difference. As the animal moved further forward through the gap, it waved its extended leg up to around the body coronal plane (average leg height = 4 ± 3 mm) (Fig. 9C). Average leg height difference was higher in the roll phase than in the other three phases ($P < 0.0001$, repeated-measures ANOVA) (Fig. 9D). Both these observations demonstrated that the animal used its legs differentially in the roll phase compared with the other three phases.

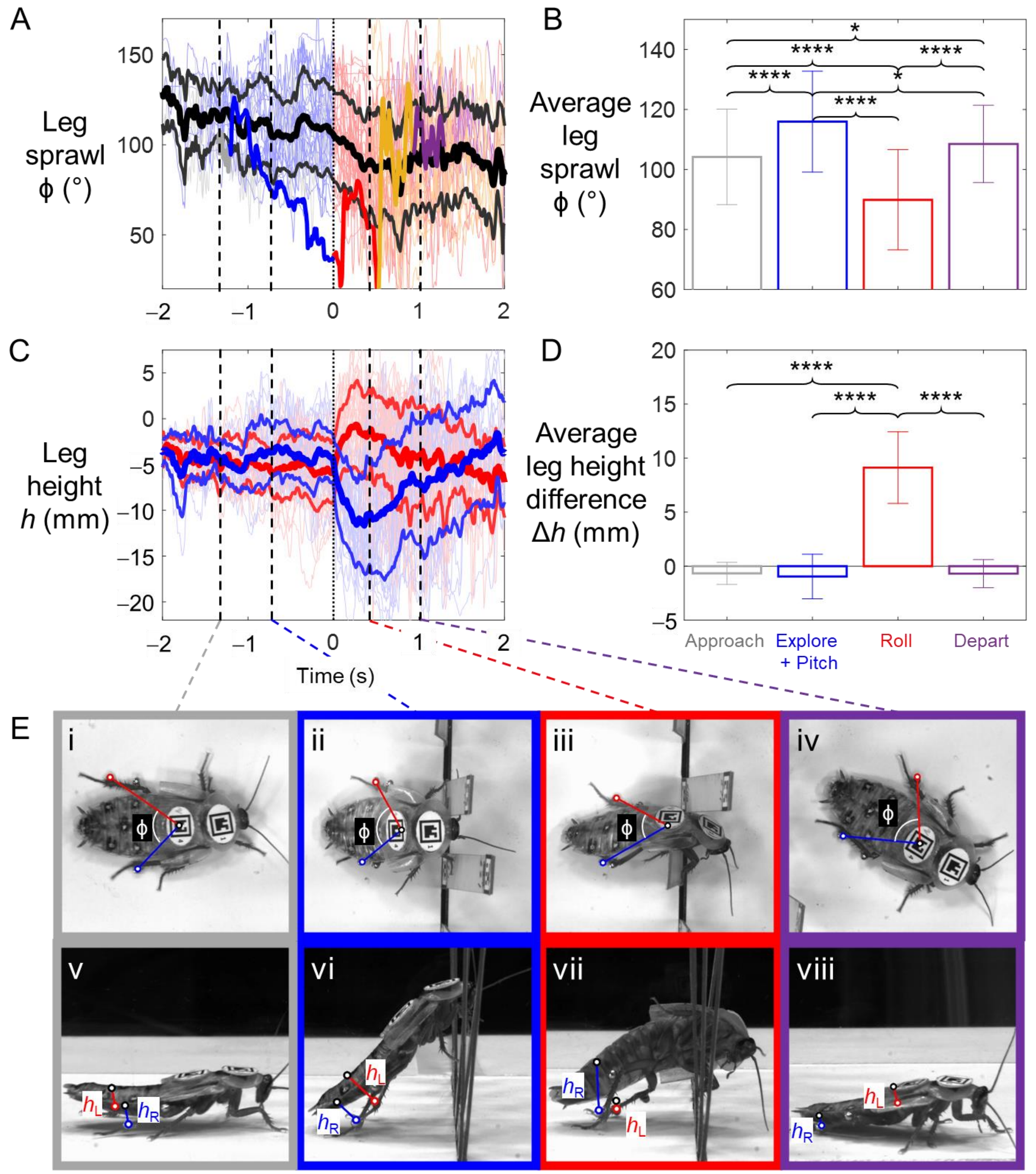

**Fig. 9. Leg adjustments statistics.** (**A**) Leg sprawl as a function of time. Thin colored curves are individual trials; thick and thin black curves are mean ± s.d. across all trials. (**B**) Leg sprawl in different phases. (**C**) Leg height of hind legs as a function of time. Blue: left leg. Red: right leg. Light-colored curves are individual trials; thick and thin black and solid colored curves are mean ± s.d. across all trials. (**D**) Leg

height difference in different phases. In (B, D), bars and error bars are means ± s.d. from temporal averaging of data in (A, C) for each phase. (**E**) Representative snapshots for each phase.

### 3.6. Refined potential energy landscape consistent with coarse landscape in previous study

The topology and evolution of the refined potential energy landscape as viewed in the pitch-roll cross-section was consistent with that in the previous study (Othayoth et al., 2020). Initially, when the animal was far from the beam, the energy landscape had a local minimum at zero pitch and roll, a basin was formed near the local minimum; as the body moved close to the beam, the basin moved along pitch direction, becoming the pitch basin, while the roll basin formed at about zero pitch and about 90° roll. Because we allowed the beam to deflect backward, the pitch basin finally went back to near zero pitch, and the roll basin finally disappeared, then the landscape became the initial landscape as the animal had traversed and moved far from the beam.

The similarity of the trend of potential energy landscape indicated that the topology of the landscape was robust against little differences in shape modeling. The landscape in the end smoothly transformed to the initial landscape showed that the beam deflection calculation was more reasonable compared to the previous study.

### 3.7. Head flexion does not facilitate pitch-to-roll transition

To test hypothesis (1), we analyzed whether adjusting head flexion can reduce the pitch-to-roll transition barrier. The transition barrier from the pitch to the roll mode with different head flexion is shown in Fig. 10A. At the average $x$ (–13.6 mm) where the animals transitioned from pitch to roll mode ,the maximum transition barrier reduction with head flexion within [–25°, 65°] was 0.003 mJ. The saved energy was small (30%) compared with the average kinetic fluctuation level. This suggested that the head adjustment did not reduce the transition barrier substantially to facilitate pitch-to-roll transition. This rejected our hypothesis that the head adjustment facilitated the mode transition by lowering the pitch-to-roll transition barrier on the potential energy landscape.

To test hypothesis (2), we analyzed whether adjusting head flexion can increase the between-gap-to-deflect transition barrier. The transition barrier from the between-gap to the deflect mode with different head flexion is shown in Fig. 10B, C. At the average $x$ (−13.6 mm) where the animal transitioned from the pitch to the roll mode, the maximal increase in transition barrier with head flexion within [−25°, 65°] was 0.012 mJ to deflect to the left and 0.008 mJ to deflect to the right. The transition barrier only increased by 10%, which required the head to hyperextend (head flexion $\beta_h = -25°$), which was rarely observed in the experiment. This suggested that the head adjustment did not increase the transition barrier substantially to prevent the animal from yawing and deflecting the beams. This rejected our hypothesis that the head adjustment facilitates the body staying within the gap.

### 3.8. Leg sprawl adjustments facilitate body rolling

To test hypotheses (3) and (4), we analyzed whether adjusting leg sprawl increases or reduces the pitch-to-roll transition barrier. The transition barrier from pitch to roll mode with different leg sprawl is shown in Fig. 10D. At $x = -20$ mm where the animals pitched against the beams (temporal average of $x$ was $-20 \pm 3$ mm in the pitch + explore phase over all trials), the transition barrier at 143° leg sprawl was larger than that at 39° leg sprawl by 0.32 mJ, which was 32 times larger than the average kinetic fluctuation (0.01 mJ). Because a larger potential energy barrier suggests higher stability, this indicated that a large leg sprawl helped the animal stay pitched up against the beams stably. At $x = -13.6$ mm where the animals transitioned from the pitch to the roll mode, the transition barrier at 39° leg sprawl was less than that at 143° leg sprawl by 0.03 mJ, which is 3 times larger than the average kinetic fluctuation level. These indicated that a small leg sprawl angle helped the animal reduce the pitch-to-roll transition barrier when rolling. Together, these findings supported our hypothesis that leg sprawl adjustment facilitated pitch-to-roll transition.

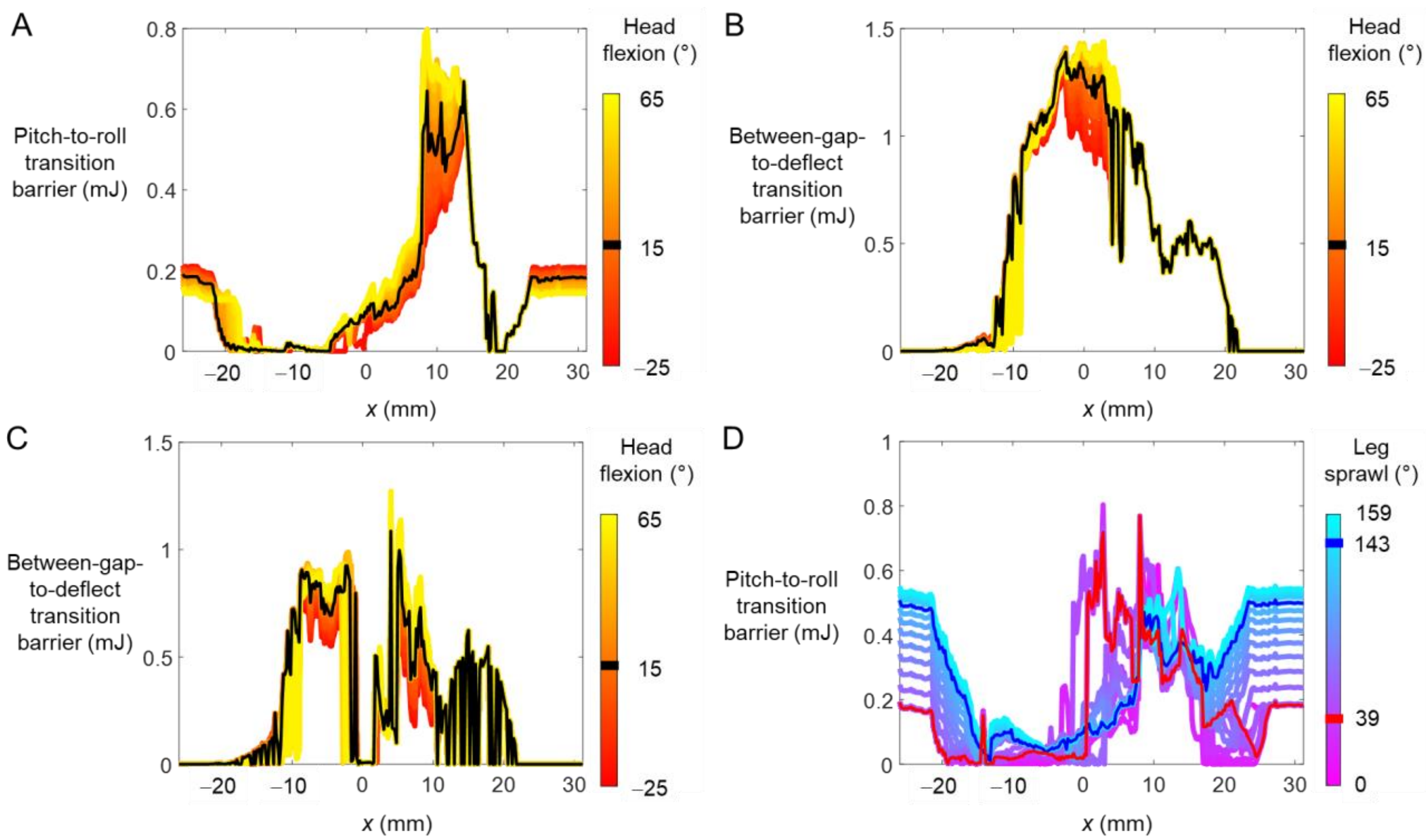

**Fig. 10. Effect of head flexion on transition barriers.** (**A**) Pitch-to-roll transition barrier as a function of *x* for different head flexion. (**B**, **C**) Roll-to-deflect transition barrier as a function of *x* under for different head flexion. (**B**) deflect to the left. (**C**) deflect to the right. In (A, B, C), black curves show transition barrier at average animal head flexion $\beta_h = 15°$. (**D**) Pitch-to-roll transition barrier as a function of *x* for different leg sprawl angles. Red and blue curves show transition barrier at leg sprawl $\phi = 39°$ and $143°$, respectively.

## 4. Discussion

As a first step to understand how animals actively control physical interaction with complex 3-D terrain to transition between locomotor modes, we quantified active adjustments by the discoid cockroach to make the pitch-to-roll transition while to traverse beam obstacles. The major adjustments included body rotations (Fig. 6), head flexion (Fig. 7), abdomen flexion (Fig. 8), and differential hind leg use (Fig. 9). Because it is strenuous to traverse the stiff beams by pushing across (the pitch mode), the animal likely made these active adjustments to facilitate transitioning to the less strenuous roll mode. Below we discuss the likely function of each adjustment and suggest future directions.

### 4.1. Role of head flexion

To find the function of active head flexion in both the explore + pitch and roll phases, we first hypothesized that by changing the overall body shape, it (1) lowered the pitch-to-roll transition barrier and (2) increased the between-gap-to-deflect transition barrier. However, we found that the head flexion did not change the transition barrier substantially in both cases. Therefore, we rejected these two hypotheses (Sec. 3.7). We speculate that this is because the cockroach's head is small and relatively more spherical (thickness/length = 0.66) compared to the thorax (0.30) and abdomen (0.21), so its orientation does not change overall body shape substantially, resulting in too small of a change in the potential energy landscape.

We speculate that the animal flexed its head to sense obstacle properties in the explore + pitch phase. Groups of campaniform sensilla and sensory hairs embedded in the cockroach's pronotum can sense the magnitude, direction, and position of the terrain reaction force (Delcomyn et al., 1996). The chordotonal organ in the animal's neck can detect the forces pushing against the head (Field and Matheson, 1998; Tuthill and Wilson, 2016). These could help the animal estimate the obstacle's physical properties (stiffness, surface friction coefficient, etc.) (Xuan et al., 2021) and guide its active adjustments to better traverse. We speculate that the occasional dynamically changing head flexion is a form of active sensing (Cellini et al., 2021; Towal, 2006). We observed that the animal seemed to flex the head upward in order to find the top end of the stiff beam obstacles (to initiate climbing) and flexed the head downward in order to find a gap to move into.

In addition, we speculate that the animal flexed its head in the later part of the roll phase (after the center of mass has passed the beams) to help its fore legs reach the ground to help propel forward, while its middle and hind legs were still interacting with the beams and were likely less effective at generating propulsion within the narrow gap. This is similar to cockroaches flexing the head to help fore legs reach the top surface when climbing a large step (Ritzmann et al., 2005).

### 4.2. Role of abdomen flexion

We speculate that the animal flexed its abdomen frequently in the roll phase to generate kinetic energy fluctuation to break resistive frictional and interlocking contact as it pushed through the beams when rolled. Because the beam gap was narrow and barely larger than the animal body thickness (average body thickness/gap width = 73%, not including legs), the animal had to retract its legs closer to the body to fit them within the gap. This made it difficult to generate thrust force from the legs. Meanwhile, the spines and other asperities on the thorax, abdomen, and legs added resistance, similar to a cockroach crawling in a confined space (Jayaram and Full, 2016). This kinetic energy fluctuation from abdomen flexion helps the animal become unstuck and thus facilitates traversal.

**4.3. Role of leg adjustments**

We speculated that the animal spread its hind legs outward to stabilize the pitched-up body against the beams in the pitch + explore phase and it tucked the hind legs inward to facilitate rolling in the roll phase. Geometrically, the stable support polygon of the animal was formed among contacts between its hind feet and the ground and contacts between its head or thorax and the beams. In the explore + pitch phase, the animal's hind legs sprawled out widely, with a large distance between the hind feet touching the ground, which increased the animal's roll stability (Fig. S3A). In the roll phase, as the animal tucked its legs inward, the support polygon shrank and roll margin of stability reduced (Fig. S3B), and the roll stability reduced. These effects can be quantified by the potential energy landscape. The potential energy barrier from the pitch to the roll basin measures the difficulty to transition. In the explore + pitch phase, a large leg sprawl increased the transition barrier, which helped the animal stay pitched up against the beams. During pitch-to-roll transition, a small leg sprawl reduced the transition barrier and made rolling easier.

Aside from leg sprawl, the differential leg uses also played an important role. We speculated that the animal extended one hind leg and retracted the other in the roll phase to generate a roll torque and it retracted both hind legs in the later part of the roll phase to reduce the resistance on the legs from the beams.

**4.4. Role of body flexibility**

These results further suggested that the flexibility from multiple flexible body parts and joints of the entire animal also contributes to reducing the efforts of mode transition and traversal (Allen et al., 2003; Jayaram and Full, 2016). The animal's body is more flexible and compliant than our model using rigid bodies. This will likely result in a smaller beam flexion than estimated by the model and reduce the transition barrier. Soft, segmented exoskeleton structure likely also reduced interlocking and frictional resistance and facilitated traversal.

**4.5. Likely involvement of sensory feedback control**

Studies on insects negotiating large obstacles have revealed that the changes in kinematics are modulated by sensory inputs. For example, when climbing large stairs, stick insects switch from using long to short steps when they have sensed a lack of substrate engagement (Theunissen and Dürr, 2013); when climbing a large step, cockroaches flex the head to help fore legs reach the top step surface when its head has sensed that it has risen above the step (Ritzmann et al., 2005). During traversal of cluttered beams in our study, the cockroach's active adjustment to make the pitch-to-roll transition is almost certainly also driven by sensory input of its environment-interaction. During traversal, it took the animal an average time of 1.3 ± 0.7 s to explore and pitch up against the beams before rolling occurred. This is well above the ~100 ms that cockroaches need to complete a feedback control loop (6-40 ms for the sensory delay (Ritzmann et al., 2012) and 47 ms for neuromuscular delay (Sponberg and Full, 2008)). We posit that the animal sensed the terrain and used this information to determine that pushing across is too strenuous and guide transitioning to the less strenuous roll mode. We have recently explored the feasibility of this in a simple simulated robotic physical model (Xuan et al., 2021).

**4.6. Future work**

Future work should verify the speculations of the mechanisms how each kind of adjustment facilitates the mode transition. (1) To understand how the flexing head facilitates sensing and how to take advantage of this, we can build a robot that is capable of the functions of head and abdomen flexing and force sensing (Wang et al., 2021) and challenge it to traverse beams while flexing its head. (2) To check whether the abdomen flexion helps the animal become unstuck, we can build a simple robot with a flexing abdomen or tail to test if the flexing of the massive lateral part helps in beams traversal. (3) To test whether active leg adjustments indeed generates a roll torque, we can add separated force plates (Goldman et al., 2006) to the ground in front of the beams to sense the ground reaction force on each foot.

To further understand the neural mechanisms involved in such cluttered large obstacle traversal, we should measure the animal's sensory neural signals (Mongeau et al., 2015; Ritzmann et al., 2012) and muscle activity (electromyogram) (Sponberg and Full, 2008; Watson et al., 2002) and alter motor activation signal to change active adjustments and test their effect on the body dynamics (Sponberg et al., 2011a; Sponberg et al., 2011b). A first challenge is to identify what sensors are involved in these cluttered adjustments for large obstacle traversal. Like other insects, cockroaches should have many sensors that can obtain information about the terrain (Delcomyn et al., 1996; Harley et al., 2009; Tuthill and Azim, 2018; Tuthill and Wilson, 2016), including: (1) visually observe the geometry of the terrain. (2) use exteroceptors like tactile hairs to sense the position of an object; (3) proprioceptors to sense relative position/velocity between joints to infer object position; and (4) campaniform sensilla and chordotonal organs to detect force and torque exerted on exoskeleton and joints. A first step to identify the relevant sensing modalities is to disabling some of these sensing sources, such as like blinding the eyes (Spirito and Mushrush, 1979) and disabling the campaniform sensilla (Pearson and Iles, 1973) and observe changes in locomotor behavior and performance. Based on animal observations, computational modeling of neural control (such as in (Schilling and Cruse, 2020)) may be fruitful for understanding feedback principles governing body and appendage adjustments to traverse cluttered large obstacles. In addition, it may be interesting to study whether animals perform active sensing (Krakauer et al., 2017; Mongeau et al., 2013; Moss et al., 2006; Nelson and MacIver, 2006; Okada et al., 2002; Schütz and Dürr, 2011; Stamper et al., 2012) in the less-considered modality of contact force sensing.

Our case study illustrated how to use fine-grained potential energy landscape modeling to understand locomotor-terrain interaction that involves active adjustments of higher dimensions, which may lead to discovery of attractive basins that result in distinct nuanced locomotor modes (see discussion in (Othayoth et al., 2020)). Fundamental insights from our study may also guide the development of multi-legged robots that can traverse cluttered terrain like forest floor for environmental monitoring (Matsuno et al., 1998), earthquake rubble for search and rescue (Freitas et al., 2011), and Martian rocks for planetary exploration (Matthies et al., 2007).

**Acknowledgments**

We thank Yuanfeng Han and Qiyuan Fu for help with imaging setup, Yuanfeng Han and Qihan Xuan for discussion, and Xiao Yu for help with animal care and manually correcting the marker tracking. This research was funded by an Arnold & Mabel Beckman Foundation Beckman Young Investigator Award, a Burroughs Wellcome Fund Career Award at the Scientific Interface, and the Johns Hopkins University Whiting School of Engineering start-up funds to C.L. Y.W. was partially supported by funds for an undergraduate research internship from the Department of Mechanical Engineering of Tsinghua University.

**Competing interests**

The authors declare no competing interest.

**Author contributions**

Conceptualization: C.L.; Methodology: Y.W., R.O., C.L.; Software: Y.W, R.O.; Validation: Y.W., C.L.; Formal analysis: Y.W.; Investigation: Y.W.; Resources: Y.W., R.O., C.L.; Data curation: Y.W.; Writing - original draft: Y.W.; Writing - review & editing: Y.W., R.O., C.L.; Visualization: Y.W.; Supervision: C.L.; Project administration: C.L.; Funding acquisition: C.L.

**Supplementary Material**

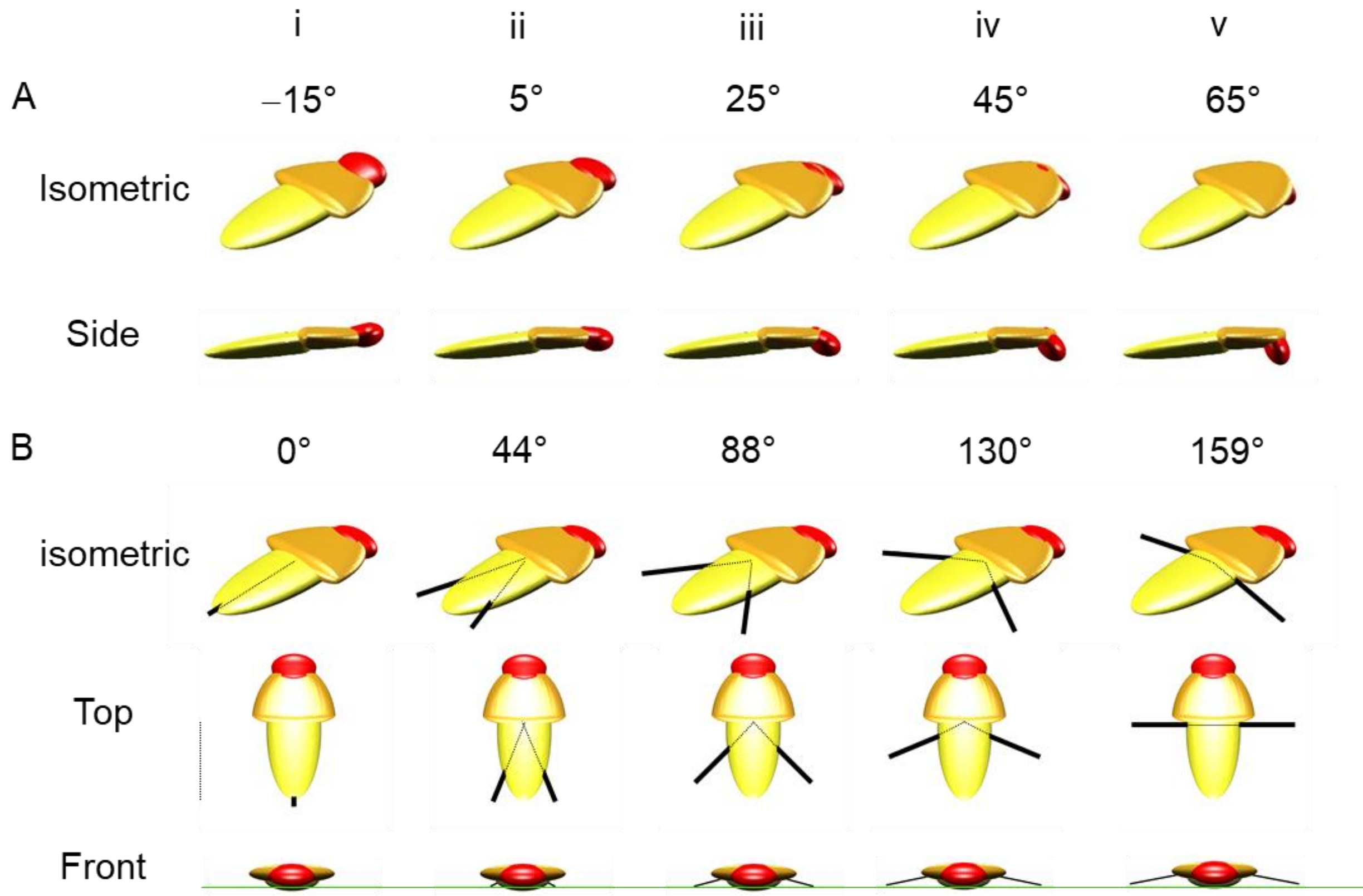

**Fig. S1. Example variation of head flexion and hind leg sprawl to test the use of head and leg adjustments in potential energy landscape model.** (**A**) Head flexion. (**B**) Leg sprawl. Green line in (**B**) side view is leg height = −5 mm.

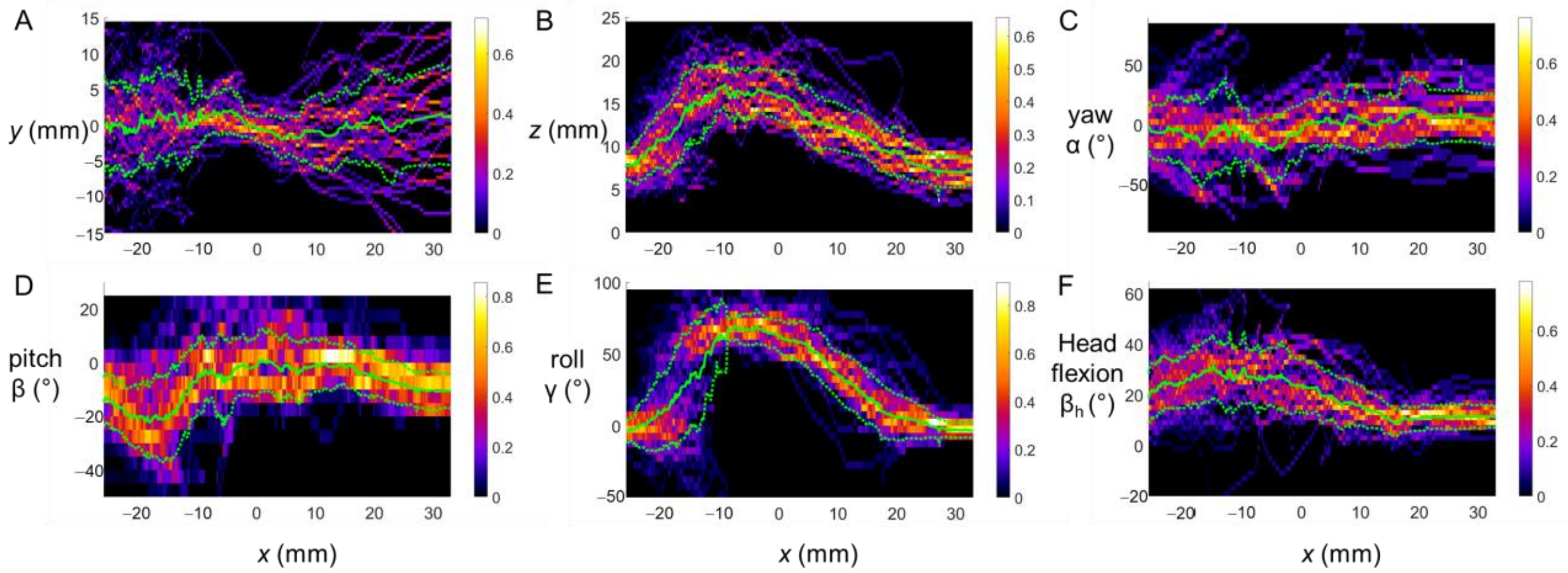

**Fig. S2. Average trajectory as a function of forwarding position *x*.** (**A**) Lateral position *y*. (**B**) Vertical position *z*. (**C**) Yaw $\alpha$. (**D**) Pitch $\beta$. (**E**) Roll $\gamma$. (**F**) Head flexion $\beta_h$. Solid and dashed green curves are mean ± s.d. from averaging data of all trials.

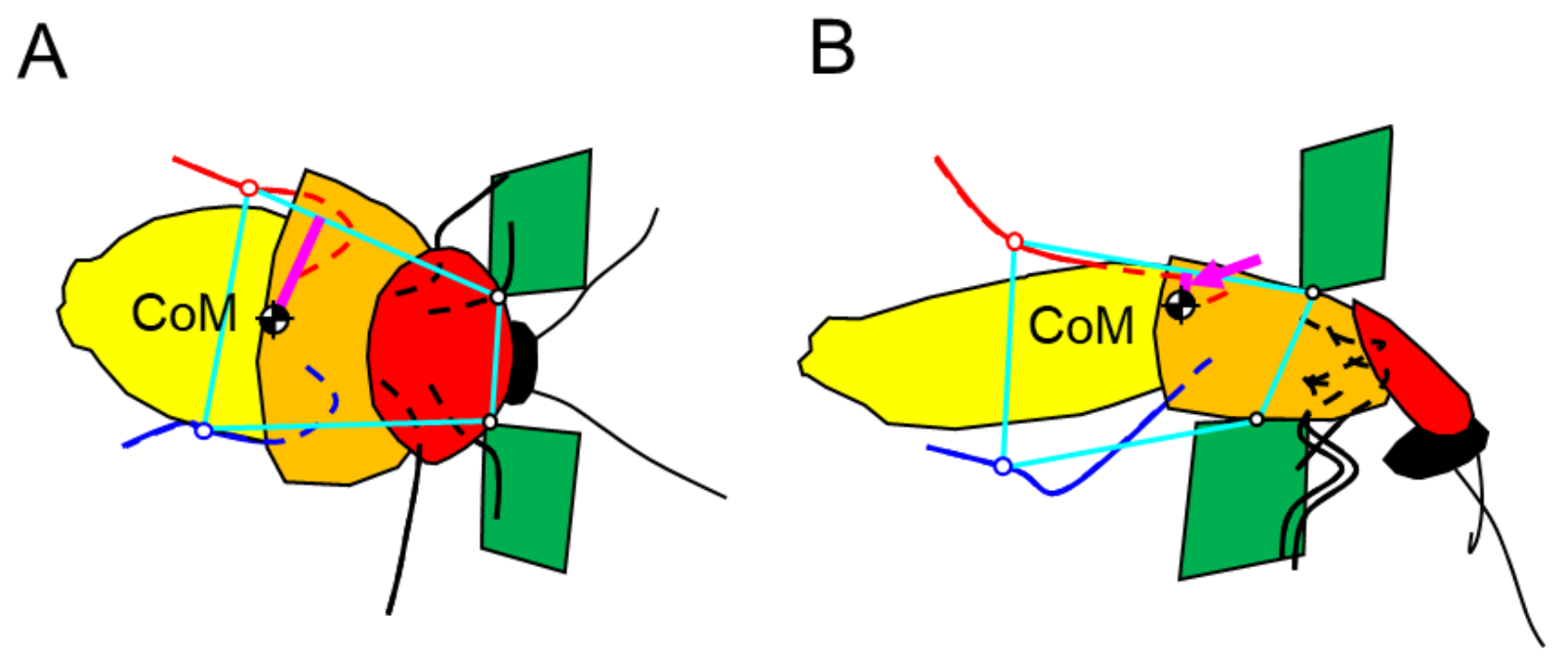

**Fig. S3. Representative support polygon evolution during pitch-to-roll transition.** (**A**) Explore + pitch phases. (**B**) Roll phase. Cyan closed shapes show support polygons, and magenta lines show the distance from center of mass (CoM) to nearest lateral edge of the support polygon, which measures roll stability. In (**B**), the distance is small and a magenta arrow is drawn.

**Table S1. Frequently used averaged variables and range**

| Variable | Time range | Measured value | Used value | First used in Sec. |
|---|---|---|---|---|
| Forward position *x* | Pitch-to-roll transition | −13.6 ± 4.4 mm | -13.6 mm | 2.9 |
| Maximum leg length | Explore + pitch and roll phase | 27 ± 2 mm | 27 mm | 2.9 |
| TA abdomen flexion $\beta_a$ | Approach phase | 7° ± 4° | 7° | 2.10 |
| Head flexion range $\beta_h$ | Whole trial | [−24°, 64°] | - | 2.12 |
| TA head flexion $\beta_h$ | Approach phase | 15° ± 4° | 15° | 2.12 |
| TA leg height | Explore + pitch phase | −5 ± 3 mm | -5 mm | 2.13 |
| Maximum leg sprawl $\phi$ | Explore + pitch phase | 140° ± 18° | 143° | 2.13 |
| Minimum leg sprawl $\phi$ | Explore + pitch phase | 40° ± 18° | 39° | 2.13 |

| Kinetic energy fluctuation | Explore + pitch, roll and land phase | 0.01 ± 0.01 mJ | 0.01 mJ | 3.1 |
|---|---|---|---|---|
| TA forward position $x$ | Explore + pitch phase | –20 ± 3 mm | -20 mm | 3.8 |

TA: temporal averaged

**Supplementary Movies**

**Movie 1.** Head flexion. Top: zoomed top (left) and side (right) views. White points with red, magenta, and green edges are origin of body frame, origin of head frame, and middle point of thorax-head joint, respectively. Solid and dotted arrows show $+x$ and $+x'$ direction of body (red) and head (magenta) frames, respectively. Head flexion is the angle between these two directions. Bottom left: isometric view. Bottom right: head flexion as a function of time.

**Movie 2.** Abdomen flexion. Top: zoomed top (left) and side (right) views. White points with red, magenta, and green edges are origin of body frame, origin of abdomen frame, and middle point of the thorax-abdomen joint, respectively. Solid and dotted arrows show $+x$ and $+x'$ direction of body (red) and abdomen (magenta) frames, respectively. Abdomen flexion is the angle between these two directions. Bottom left: isometric view. (**D**) abdomen flexion as a function of time.

**Movie 3.** Leg adjustments. Top left: zoomed top view. White points with red, blue, and black edges are the left and right tibia-tarsal joints and origin of body frame, respectively. Leg sprawl is the angle between the solid red and blue lines. Top right: zoomed side view. White points with red, blue, and black edges are tibia-tarsal joints and their projections to body coronal plane, respectively. Leg height of left and right hind legs is the length of the red and blue lines, respectively. Bottom left: isometric view. Bottom right: Leg sprawl (top) and leg height (bottom) as a function of time. Red and blue are for left and right hind legs, respectively.